\documentclass[final]{iopart}
\usepackage[final]{graphicx}
\usepackage{epsfig}

\begin{document}

\title{Revealing the degree of magnetic frustration by non-magnetic impurities}

\author{C.-C. Chen$^{1,2}$, R. Applegate$^{3}$, B. Moritz$^{1,4}$, T. P. Devereaux$^{1,5}$, R. R. P. Singh$^{3}$}
\address{$^1$SIMES, SLAC National Accelerator Laboratory, Menlo Park, California 94025, USA}
\address{$^2$Department of Physics, Stanford University, Stanford, California 94305, USA}
\address{$^3$Department of Physics, University of California, Davis, California 95616, USA}
\address{$^4$Department of Physics and Astrophysics, University of North Dakota, Grand Forks, North Dakota 58202, USA}
\address{$^5$Geballe Laboratory for Advanced Materials, Stanford University, Stanford, California 94305, USA}

\begin{abstract}
Imaging the magnetic fields around a non-magnetic impurity can provide a clear benchmark for quantifying the degree of magnetic frustration. Focusing on the strongly frustrated $J_1$-$J_2$ model and the spatially anisotropic $J_{1a}$-$J_{1b}$-$J_2$ model, very distinct low energy behaviors reflect different levels of magnetic frustration. In the $J_1$-$J_2$ model, bound magnons appear trapped near the impurity in the ground state and strongly reduce the ordered moments for sites proximal to the impurity. In contrast, local moments in the $J_{1a}$-$J_{1b}$-$J_2$ model are enhanced on the impurity neighboring sites. These theoretical predictions can be probed by experiments such as nuclear magnetic resonance and scanning tunneling microscopy, and the results can elucidate the role of frustration in antiferromagnets and help narrow the possible models to understand magnetism in the iron pnictdies.
\end{abstract}

\pacs{74.25.nj, 74.40.Kb, 74.70.Xa, 75.30.Hx}

\date{\today}
\maketitle

One of the major thrusts in condensed matter physics concerns the interplay of strong frustration and quantum fluctuations in magnetic systems~\cite{Balents}. These systems can exhibit (a) novel phases of matter lacking conventional order, (b) a high degree of residual entropy at low temperatures with local constraints leading to emergent gauge fields~\cite{gauge_field}, and (c) new types of particles with unusual quantum numbers and dynamics not captured within the established framework of quasi-particles~\cite{Kivelson}. But, how can such magnetic frustration be recognized and quantified experimentally?

The ratio of the Curie-Weiss temperature to the magnetic ordering temperature has served as an important metric~\cite{Ramirez}, though it cannot distinguish between the role of fluctuations due to reduced dimensionality and that due to frustration. The spin-wave spectrum can at times lead uniquely to the underlying magnetic interactions, though there are cases where this also can be ambiguous. In some classical systems, such as spin-ice, the residual entropy can be measured directly by specific heat~\cite{spin=ice}. Various ratios associated with the peak in the uniform susceptibility also can quantify frustration~\cite{coldea-mckenzie}. However, thermodynamic measurements are not easy to interpret when other low energy degrees of freedom, besides spins, are present. Since frustration is a local property, measurements which can infer it locally would be useful.

Studying antiferromagnetism, magnetic frustration, and quantum criticality in the iron pnictides is crucial for a comprehensive understanding of their superconductivity. This also can lead to the development of a broader framework for unconventional superconductivity in materials such as cuprates, ruthenates, heavy fermions, and organic superconductors~\cite{Doug}. \emph{In this letter, we explore how imaging the magnetic fields around a non-magnetic impurity can provide a clear benchmark for quantifying the degree of magnetic frustration}. While the pnictides provide the context for this study, it should have a broader applicability in frustrated spin systems~\cite{Carretta}.

Theories for the collinear antiferromagnetic (AF) phase in iron pnictides vary greatly in the importance of weak versus strong coupling~\cite{Kuroki, Dagotto, Chubukov, Mazin, Kou}, and in the amount of magnetic frustration and proximity to a quantum critical point (QCP)~\cite{Yildirim, QCP_PNAS}. There are various reasons why a local moment perspective is still relevant in these systems. Parametrization of the magnetic behavior by \emph{effective} local moment models,  even for metals, should remain valid as long as there are long-lived magnon-like excitations~\cite{Thalmeier, D_Johnston}. Most importantly, however, is that \emph{local moment models are still being widely used in experiments to explain magnetic properties in these materials}. These include the spatially anisotropic $J_{1a}$-$J_{1b}$-$J_2$ model~\cite{J1abJ2_1, J1abJ2_2, Neutron_Dai}, where the coupling is ferromagnetic (FM) in one direction and AF in the other, and the strongly frustrated $J_1$-$J_2$ model~\cite{J1J2_1, J1J2_2, J1J2_3}, where collinear AF order arises as a result of a subtle selection via order by disorder~\cite{CCL}.

Neutron scattering on CaFe$_2$As$_2$ indicates that the spin wave energy is a maximum at momentum transfer $(\pi,\pi)$~\cite{Neutron_Dai}, which favors the $J_{1a}$-$J_{1b}$-$J_2$ scenario. However, the results have been disputed~\cite{Neutron_Diallo} and alternative experiments would be useful. A particularly good example of competing models is the iron-chalcogenide superconductors, Fe$_{1+y}$Te$_{1-x}$Se$_{x}$, where the ($\pi/2,\pi/2$) AF order can be obtained by either invoking strong frustration in a $J_1$-$J_2$-$J_3$ model~\cite{Hu_11} or utilizing a model with strong spatial anisotropy arising from orbital order~\cite{Ashvin_11}. Both models can lead to virtually indistinguishable spin-wave spectra. Hence, further experiments are necessary to clarify the nature of the magnetic phase.

Leaving aside the question of where such strong in-plane anisotropy can come from~\cite{Kruger, spin-orbital, Ku, Lv_1, Lv_2, new_rapid}, one can ask whether experiments can directly sense the degree of frustration in the system. In the pnictides, one way to distinguish models with different levels of frustration is to study the response to non-magnetic impurities~\cite{RMP}, such as those caused by Zn substitution~\cite{DSingh}. Such impurities can suppress or enhance antiferromagnetism in their neighborhood, and lead to bound magnon states with different degrees of anisotropic perturbations.

Using a $T$-matrix approach within linear spin-wave theory~\cite{Bulut}, in the present study we find that the static and dynamic perturbation of collinear antiferromagnetism near a non-magnetic impurity are markedly different in the two models. In the frustrated $J_1$-$J_2$ model, a non-magnetic impurity strongly reduces its neighboring local moments and overturned dynamical spins appear close to zero energy, rendering non-trivial physics in proximity to the QCP. In contrast, the $J_{1a}$-$J_{1b}$-$J_2$ model produces enhanced local moments on the impurity neighboring sites, a behavior similar to the Heisenberg model~\cite{Bulut}. In both models, the disturbance on the magnon local density of states (LDOS) heals quickly along the AF direction but is extended in the FM direction. This leads to spin fluctuations with an anisotropic stripe pattern.

\begin{figure}[t]
\begin{center}
\includegraphics[width=4in]{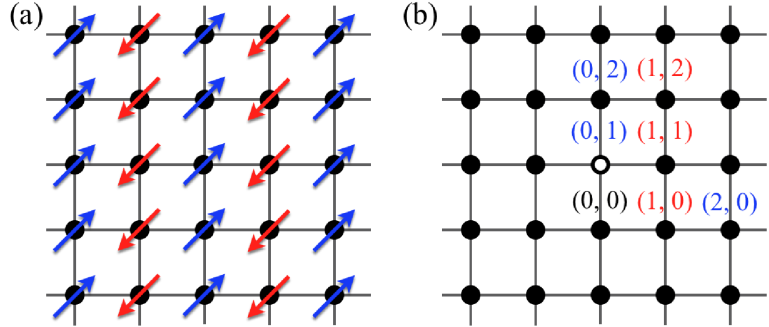}
\end{center}
\caption{(Color online) (a) Schematic for the collinear AF state on a 2D square lattice. (b) The position labelings for selected sites. The impurity is placed at (0,0).
}\label{fig:lattice}
\end{figure}

We consider a spin Hamiltonian on a 2D square lattice
\begin{equation}
H_0= \sum_r \left( J_{1a} \mathbf{S}_r\cdot \mathbf{S}_{r+\hat{x}}+J_{1b} \mathbf{S}_r\cdot \mathbf{S}_{r+\hat{y}}\right) + \sum_{\ll i,j \gg} \frac{J_2}{2} \mathbf{S}_i \cdot \mathbf{S}_j,
\end{equation}
where $\ll ...\gg$ represents a second nearest-neighbor sum. When $J_{1b}=J_{1a}$ the Hamiltonian reduces to the $J_1$-$J_2$ model. For a collinear AF order, we introduce the real space Holstein-Primakoff bosons in the linear spin wave approximation: $S^z_i =  S-n_i\, S^\dagger_i=\sqrt{2S} a_i$, and $S^z_j = -S+n_j, \,S^\dagger_j = \sqrt{2S} b^\dagger_j$. We use the notation that $i$ belongs to sub-lattice A (spin up), and $j$ belongs to sub-lattice B (spin down) [Fig. 1(a)]. A standard Bogoliubov-de Gennes transformation diagonalizes $H_0$ with eigen-energies $E_k=J_2SZ\sqrt{A^2_k-B^2_k}$, where $k$ is defined over the reduced (magnetic) Brillouin zone. Here $A_k\equiv 1+\alpha+\beta(\cos k_y-1)$, and $B_k\equiv\cos k_x (\cos k_y+\alpha)$. $Z$ is the coordination number, and $\alpha\equiv J_{1a}/2J_2$, $\beta\equiv J_{1b}/2J_2$.

For the $J_1$-$J_2$ model, we focus on $J_2=J_1$~\cite{Hayden}. In the collinear AF state, the results for various $J_2/J_1$ ratios are qualitatively the same, except for $J_2/J_1$ very close to the QCP as we discuss below. For the $J_{1a}$-$J_{1b}$-$J_2$ model, we use $J_{1b}=-0.1J_{1a}$ and $J_2=0.4J_{1a}$, appropriate for CaFe$_2$As$_2$~\cite{Neutron_Dai}. In the latter case the system is unfrustrated even if $(J_{1a}+J_{1b})$ is comparable to $2J_2$.

The effects of a single non-magnetic impurity are introduced by removing the exchange interactions between the impurity and its first and second nearest neighbor sites. The impurity problem is then solved conveniently by defining the Green's function matrix:
\begin{eqnarray}
\hat{G_{r,r'}}(t)&=&\left(\begin{array}{cc}
-i\langle T a^\dagger_i (t) a_{i'}(0)\rangle & -i\langle T a^\dagger_i (t) b^\dagger_{j'}(0)\rangle  \\
-i\langle T b_j(t) a_{i'}(0)\rangle  & -i\langle T b_j (t) b^\dagger_{j'}(0)\rangle  \\ \end{array}\right).
\end{eqnarray}
The bare Green's functions in momentum and frequency space are
\begin{eqnarray}
(\hat{G}^0_{k,k'})_{\stackrel{11}{^{(22)}}}(\omega)&=&\delta_{k,k'} \lim_{\eta\rightarrow 0^+} \frac{\mp\omega+J_2SZ A_k}{\omega^2-E^2_k+i\eta},
\nonumber\\
(\hat{G}^0_{k,k'})_{\stackrel{12}{^{(21)}}}(\omega)&=&\delta_{k,k'} \lim_{\eta\rightarrow 0^+} \frac{-J_2SZ B_k} {\omega^2-E^2_k+i\eta}.
\end{eqnarray}
The dressed Green's functions $\hat G$ in the presence of an impurity can be calculated \emph{exactly in the thermodynamic limit} (within spin wave approximation) by the $T$-matrix formalism~\cite{Bulut}. This theoretical treatment has ensured that free bosons associated with the impurity site never hop and thus avoids spurious zero modes~\cite{Chernyshev}. The physical quantities of interest include the magnetic moments, as well as the magnon LDOS, which can be obtained from the imaginary part of the Green's function~\cite{Bulut}.

\begin{figure}[t]
\begin{center}
\includegraphics[width=4in]{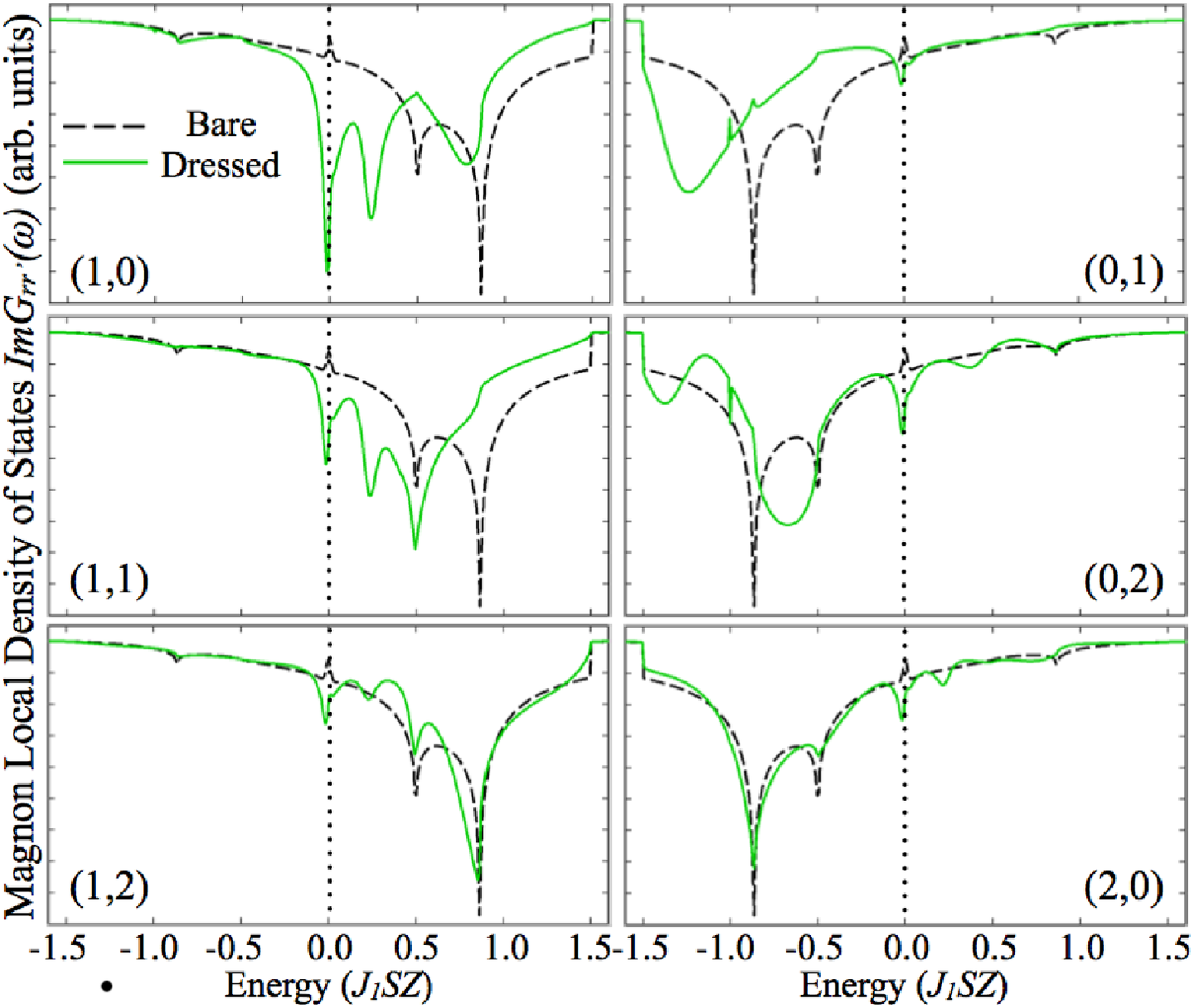}
\end{center}
\caption{(Color online) $J_1$-$J_2$ model: The imaginary parts of the Green's functions related to the magnon LDOS at selected positions. The magnetic disturbance heals quickly along the AF direction but extends more in the FM direction.
}\label{fig:J1J2_LDOS}
\end{figure}

\begin{figure}[t!]
\begin{center}
\includegraphics[width=4in]{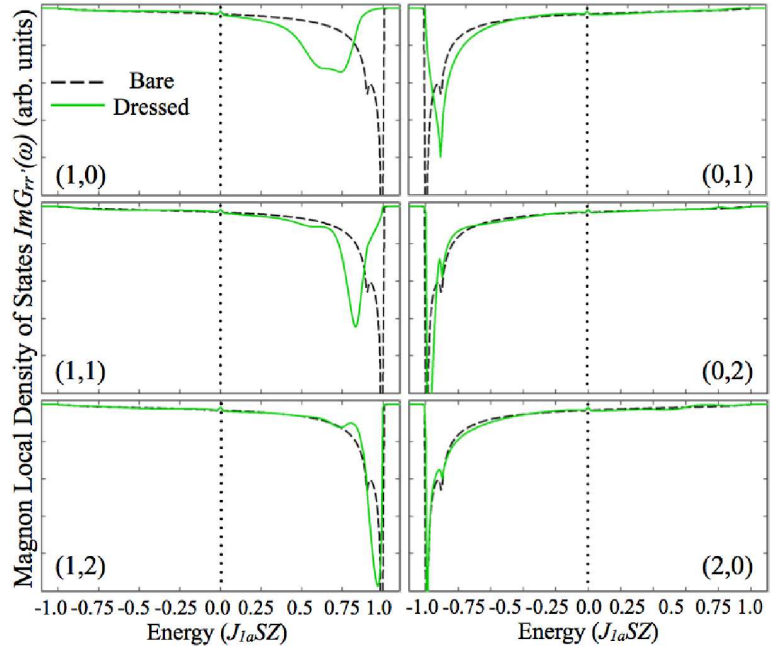}
\end{center}
\caption{(Color online) $J_{1a}$-$J_{1b}$-$J_2$ model: The imaginary parts of the Green's functions related to the magnon LDOS at selected positions. The magnetic disturbance heals quickly along the AF direction but extends more in the FM direction.
}\label{fig:J1abJ2_LDOS}
\end{figure}

One of our most interesting findings is the presence of overturned dynamical spins near the impurity in the ground state for the $J_1$-$J_2$ model, but not for the $J_{1a}$-$J_{1b}$-$J_2$ models. In the $J_1$-$J_2$ model, the magnon LDOS for spin sub-lattice B [left panels of Fig. 2] shows more low energy features, and localized magnon excitations appear near the impurity at finite, positive energies~\footnote{There are artificial zero modes at $k=(\pi,\pi)$ in the linear spin wave spectra in the $J_1$-$J_2$ model, which is known to be gapped by quantum fluctuations or spin anisotropy~\cite{uhrig, ryan}. By making $J_{1a}$ and $J_{1b}$ unequal, we have verified that our results change only slightly when a spin-wave gap at $k=(\pi,\pi)$ is gradually turned on.}.
In contrast, the LDOS in the $J_{1a}$-$J_{1b}$-$J_2$ model shows features only at relatively high energies close to the van Hove singularity of the homogeneous host system, as shown in Fig. \ref{fig:J1abJ2_LDOS}. In both models, the disturbance on the magnon LDOS heals quickly along the AF direction but extends more in the FM direction.

\begin{figure}[t]
\begin{center}
\includegraphics[width=4in]{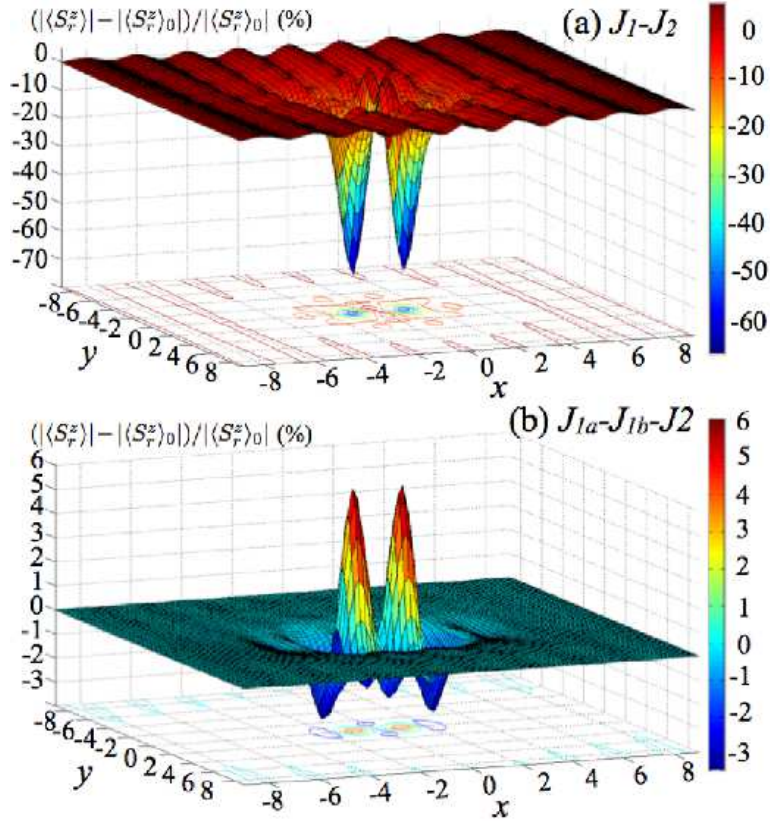}
\end{center}
\caption{(Color online) Changes in the magnetic moment for (a) $J_1$-$J_2$ model and (b) $J_{1a}$-$J_{1b}$-$J_2$ model.  A spline interpolation is used for both plots. The moment at site $r=(1,0)$ $\vert \langle S^{z}_{r=(1,0)}\rangle\vert$ is strongly reduced in the $J_1$-$J_2$ model indicating a stronger quantum fluctuation, while for the $J_{1a}$-$J_{1b}$-$J_2$ model $\vert \langle S^{z}_{r=(1,0)}\rangle\vert$ is slightly enhanced. The change in magnitude of the moments in the $J_{1a}$-$J_{1b}$-$J_2$ model is $\sim$ 10$\%$, but this change in the $J_1$-$J_2$ model is more substantial. Both models exhibit spin fluctuations with an anisotropic stripe pattern.}\label{fig:moment_surfc}
\end{figure}

We also find a strong reduction in the magnetic moments near the impurity sites in the frustrated model, but not in the unfrustrated case. Figure \ref{fig:moment_surfc} shows the change in magnitude of the magnetic moments ($\vert \langle S^z_r\rangle \vert-\vert \langle S^{z}_r\rangle_0\vert)/\vert \langle S^{z}_r\rangle_0\vert$ for the two models. In the frustrated $J_1$-$J_2$ model, at site $r=(1,0)$ (belonging to a different spin sub-lattice from the impurity), the magnitude of the moment strongly decreases, indicating enhanced quantum fluctuations [Fig. 4(a)]. On the other hand, in the unfrustrated $J_{1a}$-$J_{1b}$-$J_2$ model the magnitude of the moment increases [Fig. 4(b)], a behavior more similar to the Heisenberg model~\cite{Bulut}. Moreover, while the change in magnitude of the moments in the $J_{1a}$-$J_{1b}$-$J_2$ model is $\sim$ 10$\%$, this change in the $J_1$-$J_2$ model is more substantial. In both models, the change of the moments is damped quickly away from the impurity by a few lattice sites, but exhibits spin fluctuations of an anisotropic stripe pattern. Density functional calculations on cobalt doped pnictides also found a similar behavior~\cite{Lex}. We emphasize that these fluctuations are associated with the spin variables, which must be distinguished from the low-energy charge excitations in metals.

Aside from changing the magnon LDOS and the magnetic moments, non-magnetic impurities can also result in an instability of the ground state~\cite{metlitski}. In the spin wave calculation, to sustain a collinear AF ground state in the $J_1$-$J_2$ model requires $J_2\ge 0.5J_1$. With the presence of a non-magnetic impurity, our calculations for $J_2/J_1<0.6$ indicate that the moments are drastically reduced, and the changes in moments exceed the bare values. In this case, the system no longer favors a collinear AF state and the spin wave calculation breaks down. In the pnictides, the $J_1$-$J_2$ model requires fine-tuning to the QCP to explain the reduced ordered moment~\cite{uhrig}. Therefore, a competing phase other than the collinear AF state may result upon doping by non-magnetic impurities.

In principle, the ordered local moments can be measured directly in nuclear magnetic resonance (NMR), where they result in a local magnetic field on the nucleus and cause a shift in its resonance frequency~\cite{Halperin}. Our calculations for non-magnetic impurities (such as Zn substituting Fe) indicate that the $J_1$-$J_2$ model produces a more substantial broadening of the magnetic field distribution, while this inhomogeneous broadening in the $J_{1a}$-$J_{1b}$-$J_2$ model is much smaller. On the other hand, in both cases the presence of line defects such as those coming from domain boundaries would tend to spread the distribution. These issues as well as confirmation of the anisotropic disturbance of the collinear AF state deserve further experimental attention. We last note that recent NMR measurements on Ni doped BaFe$_2$As$_2$ found that the system remains in a collinear AF state upon doping~\cite{diouguardi}. The experiments also show a considerable broadening of the field distribution. However, it is not clear to what extent Ni or Co dopants can be treated as localized non-magnetic impurities~\cite{Sawatzky, Ikeda}.

Spin-resolved scanning tunneling microscopy (STM) can potentially measure the full frequency dependence of the spin structure~\cite{SPSTM_1, SPSTM_2}. One complication is that there are other low energy (charge) degrees of freedom present besides spin. Indeed, STM measurements for low energy quasi-particles on the pnictides show unidirectional nanostructures weakly pinned by dopant atoms, strongly suggestive of broken tetragonal symmetry and in-plane anisotropy~\cite{STM_1, STM_2}. On the other hand, the spin-resolved magnetic properties may be embedded under signals from the charge degrees of freedom, and proper care is needed to separate the tunneling conductance between localized spins and itinerant charges. A spin-polarized STM with atomic resolution can filter out the magnon-like excitations, whose projection is fully spin polarized at every site, from those itinerant charge degrees of freedom, which may be unpolarized or weakly spin polarized.

In summary, we have studied the responses to a non-magnetic impurity for the $J_1$-$J_2$ and $J_{1a}$-$J_{1b}$-$J_2$ models in the collinear AF state. We found that frustration in the $J_1$-$J_2$ model results in bound magnons at very low energies with concomitant reduction of the magnetic moments of the impurity neighboring sites. In contrast, the $J_{1a}$-$J_{1b}$-$J_2$ model shows behavior more similar to the square-lattice Heisenberg model, where local moments are enhanced on the sites proximal to the impurity. We made clear experimental predictions for magnon LDOS and real space spin textures that potentially can be measured by NMR, STM or other resonance probes. The results can benchmark strong frustration and proximity to a QCP in the system under study.

We last note that the results discussed above should qualitatively carry over to other spin systems with different AF order, such as the ($\pi/2,\pi/2$) order in the Fe$_{1+y}$Te$_{1-x}$Se$_{x}$ materials. Due to differences of frustration intrinsic to the proposed models~\cite{Ashvin_11,Kruger}, the magnon LDOS and the response of the AF hosts to non-magnetic impurities can be rather different. Calculations related to this interesting extension are an area for future work.
 
\ack
The authors acknowledge discussions with A.~F. Kemper, N.~J. Curro, A.~P. Dioguardi, J.-H. Chu, I.~R. Fisher, M. Gingras, S. Raghu, and S.~A. Kivelson.  CCC, BM, and TPD are supported by the U.S. DOE under Contract No. DE-AC02-76SF00515. RA and RRPS are supported by NSF Grant No. DMR-1004231. This research used resources of NERSC, supported by U.S. DOE under Contract No. DE-AC02-05CH11231.

\section*{References}

\bibliographystyle{iopart-num}
\bibliography{Impurity_NJP_Resubmitted}

\end{document}